\documentclass[preprint,
aps,prd,nofootinbib,superscriptaddress,tightenlines]{revtex4}

\usepackage{amsmath, amsfonts, amsthm, amssymb, graphicx,comment}

\allowdisplaybreaks[3]

\def\be{\begin{equation}}
\def\ee{\end{equation}}
\def\bea{\begin{eqnarray}}
\def\eea{\end{eqnarray}}
\newcommand{\nn}{\nonumber\\}
\newcommand{\ud}{\mathrm{d}}
\def \pd {\partial}

\begin{document}

\title{Mass-Varying Massive Gravity}

\author{Qing-Guo Huang}
\email[]{huangqg@itp.ac.cn}
\affiliation{State Key Laboratory of Theoretical Physics, Institute of Theoretical Physics,
Chinese Academy of Sciences, P.O. Box 2735, Beijing 100190, China}

\author{Yun-Song Piao}
\email[]{yspiao@gucas.ac.cn} \affiliation{College of Physical
Sciences, Graduate University of Chinese Academy of Sciences,
Beijing 100049, China}

\author{Shuang-Yong Zhou}
\email[]{zhou.sy234@gmail.com}
\affiliation{School of Physics and Astronomy, University of Nottingham, Nottingham, NG7 2RD, UK}

\date{\today}

\begin{abstract}
It has recently been shown that the graviton can consistently gain
a constant mass without introducing the Boulware-Deser ghost.  We
propose a gravity model where the graviton mass is set by a scalar
field and prove that this model is free of the Boulware-Deser
ghost by analyzing its constraint system and showing that two
constraints arise. We also initiate the study of the model's
cosmic background evolution and tentatively discuss possible
cosmological implications of this model. In particular, we
consider a simple scenario where the scalar field setting the
graviton mass is identified with the inflaton and the graviton
mass evolves from a high to a low energy scale,
giving rise to the current cosmic acceleration.
\end{abstract}


\maketitle

\section{Introduction}

Although known as totally redundant degrees of freedom, gauge symmetries and general covariance are important for the modern understanding of particle physics and gravity respectively. While gauge symmetries can be straightforwardly broken to promote a massless gauge particle to a massive one, the equivalent for general covariance is more difficult to achieve \cite{Hinterbichler:2011tt}. Phenomenologically, the Standard Model gauge symmetries are spontaneously broken at low energies, where we are living, and gauge particles gain masses from the condensate of the Higgs boson. It is interesting to ask whether a similar situation applies to the graviton, with the graviton mass smaller than the bound imposed by experiments? The recently observed cosmic acceleration  may add another motive for considering this scenario (or modified gravity theories in general \cite{Clifton:2011jh}), where the dark energy scale is introduced via the graviton mass.

The linear massive spin-2 theory was discovered as early as Fierz-Pauli \cite{Fierz:1939ix}. However, the gravitational force described by this theory differs significantly from that of linearized General Relativity, for example, in the solar system. One can compute the scattering amplitude of Fierz-Pauli theory between two matter sources and find that when the mass goes to zero the amplitude generally does not reduce to the linearized General Relativity case. This is known as the vDVZ discontinuity \cite{vanDam:1970vg, Zakharov:1970cc} and seems to suggest that by simply performing experiments in localized regions one can infer that the graviton mass is mathematically zero, contradicting the expectation that forces mediated by bosons decay {\it \`{a} la} Yukawa.  As pointed out by Vainshtein \cite{Vainshtein:1972sx}, in nonlinearized Fierz-Pauli models, the gravitational force should mimic the Einstein case in conventional environments for a small mass: A key observation was that the introduction of the graviton mass has introduced a second nonlinear scale different from the one associated with the Planck scale. For example, around a spherical compact source, General Relativity is recovered within the nonlinear length scale called the  Vainshtein radius, while linear Fierz-Pauli theory applies outside the Vainshtein radius. The Vainshtein radius is very large for a small graviton mass and goes to infinity when the mass goes to zero. However, nonlinear Fierz-Pauli models are not free from problems. The most pressing one is that there is a ghost originally identified by Boulware-Deser (BD) \cite{Boulware:1973my}. Whilst a spin-2 particle should only have 5 propagating modes, nonlinear Fierz-Pauli gravity has 6 modes: The phase space of a generic 4 dimensional metric theory is 12 dimensional, as there are, after a Arnowitt-Deser-Misner (ADM) decomposition, 6 apparent dynamical components in the metric (the spatial ones); In nonlinear Fierz-Pauli gravity, there are no constraints at all, comparing to 8 constraints in the General Relativity case and 2 constraints in the linear Fierz-Pauli case. The sixth mode in nonlinear Fierz-Pauli models is responsible for the BD ghost \cite{Creminelli:2005qk}.

Although broken in massive gravity, general covariance can be re-introduced by the St\"{u}ckelberg trick, i.e., introducing extra scalar fields patterned after general coordinate transformation \cite{ArkaniHamed:2002sp}. In this way, we can easily decompose all the dynamical modes into a scalar, vector and tensor, and view the theory in the effective field theory's language, where the peculiarities of massive gravity such as the vDVZ discontinuity, Vainshtein mechanism and BD ghost can be clearly understood \cite{ArkaniHamed:2002sp}. In particular, most clearly seen in the decoupling limit, the scalar sector contains higher derivative interactions that are strongly coupled at $\Lambda_5=(m_g^4 M_P)^{1/5}$ ($m_g$ is the graviton mass.), which is rather small for a small mass. The scalar's equation of motion contains fourth order derivatives, so by a Cauchy surface dimension counting, the scalar actually propagates two modes and, according to Ostrogradski's theorem, one of the scalar modes is a ghost.

However, as recently discovered, all the $\Lambda_5$ interactions can be systematically removed and the strong coupling scale can be raised to $\Lambda_3=(m_g^2 M_P)^{1/3}$ in de Rham-Gabadadze-Tolley massive gravity \cite{deRham:2010kj, deRham:2010gu, deRham:2010ik}, eliminating the BD ghost at least in the decoupling limit. This construction is closely connected to the galileon construction, which had been studied in a different class of modified gravity models (see e.g.~\cite{Nicolis:2008in, Padilla:2010de, Padilla:2010tj}). In fact, the sixth BD mode is eliminated in the full dRGT model. This has been shown by analyzing the constraint system of the theory in the Hamiltonian formulation \cite{Hassan:2011hr, Hassan:2011ea, Hassan:2011tf, Hassan:2011vm, deRham:2010kj, deRham:2010ik}. Unlike nonlinear Fierz-Pauli massive gravity, the dRGT model maintains the Hamiltonian constraint and a secondary constraint is generated by the time conservation of the Hamiltonian constraint \cite{Hassan:2011hr, Hassan:2011ea}, confining the system to evolve in a 10 dimensional surface in the full phase space. The absence of the BD ghost has been checked from different perspectives and various further consistency issues of the dRGT model have also been examined \cite{deRham:2011rn, Hassan:2012qv, deRham:2011qq, Golovnev:2011aa, Kluson:2012wf, deRham:2011pt, deRham:2012kf, Koyama:2011yg, Chkareuli:2011te, Burrage:2011cr, Comelli:2012vz, Sbisa:2012zk}.

In massive gravity, there is a second fiducial metric, which is promoted to be dynamical in bi-metric theories. It has been shown that the bi-metric model generalized from dRGT massive gravity is also free of the BD ghost \cite{Hassan:2011zd, Hassan:2011ea}. The multi-Vierbein and multi-metric model that are free of the BD ghost have also been formulated \cite{Hassan:2012wt, Hinterbichler:2012cn}. Solutions of dRGT massive gravity and bi-metric gravity have been investigated by various authors \cite{Volkov:2011an, Volkov:2012cf, Gumrukcuoglu:2011ew, Koyama:2011xz, Nieuwenhuizen:2011sq, Comelli:2011zm, vonStrauss:2011mq, Chamseddine:2011bu, Comelli:2011wq, D'Amico:2011jj, deRham:2010tw, Berezhiani:2011mt, Gumrukcuoglu:2011zh, Crisostomi:2012db, Gratia:2012wt, DeFelice:2012mx, Kobayashi:2012fz, D'Amico:2012pi}.

In this paper, we generalize dRGT massive gravity so that the graviton mass can vary in time and space. This can be most easily done by introducing a scalar whose background value sets the graviton mass. (Promoting the mass parameter to a function of a scalar has been briefly mentioned previously \cite{D'Amico:2011jj} but without detailed discussion on its self-consistency and generic cosmology.  When this work was being completed, \cite{D'Amico:2012zv} appeared which generalizes the dRGT model with a quasi-dilaton scalar, from a rather different perspective.) This mass-varying massive gravity model is free of the BD ghost, as we will explicitly show by examining the constraint system in the Hamiltonian formulation.  We will also study the FRW cosmology of this model. In this scalar-dRGT graviton system, plenty of interesting cosmological scenarios might be possible. Among them, we will tentatively discuss a scenario where the scalar plays the role of the chaotic inflaton in the early universe and its low energy vev gives rise to dark energy.

The paper is organized as follows. In Section~\ref{themodelsec}, we describe the mass-varying massive gravity model we are considering. In Section~\ref{absencebdghost}, we prove that mass-varying massive gravity is free of the BD ghost. We will first show that the model has a Hamiltonian constraint and then prove that there is a secondary constraint generated by the Hamiltonian constraint and calculate the secondary constraint explicitly. In Section~\ref{frwcosmology}, we discuss the homogenous and isotropic evolution of the model, focusing on the case when the fiducial metric is Minkowski. In particular, we will derive the background equations of motion for the case that the dynamical and fiducial metrics can be diagonalized simultaneously, and briefly discuss a simple model that unifies inflation and dark energy. We conclude and outline possible further consistent generalizations of our model in Section~\ref{conclusion}.

\section{The model}  \label{themodelsec}

As in the Standard Model, the mass of a particle can be generated by a condensate of a scalar field.  A constant mass can be generated by a vev, while a condensate of the $\mathbf{k}=0$ mode generates a mass parameter that is a function of time. For massive gravity to have a varying mass, we may promote the quantity $m_g^2 M_P^2$, $m_g$ being the graviton mass, to be a function of a scalar field $V(\psi)$. 
General covariance is spontaneously broken unless the scalar sits at $V(\psi)=0$.  Adding a kinetic term and a potential, we will consider the mass-varying massive gravity model
\be \label{thefirstmodel}
S= \int \ud^4 x \sqrt{-g} \left[ \frac{M_P^2}{2} R + V(\psi) ( U_2 + \alpha_3 U_3 + \alpha_4 U_4)  -  \frac12 \pd_\mu \psi \pd^\mu \psi -W(\psi) \right]   .
\ee
The dRGT massive gravity \cite{deRham:2010kj} is the special case of our model when $\psi$ sits in a non-zero vev. As in dRGT gravity, $\alpha_3$ and $\alpha_4$ are free dimensionless parameters (They may also be consistently promoted to functions of $\psi$; see Discussions.), and the three graviton potentials read
\begin{align}
U_2  =  \mathcal{K}^\mu_{[\mu}\mathcal{K}^\nu_{\nu]} , \quad
U_3  = \mathcal{K}^\mu_{[\mu}\mathcal{K}^\nu_{\nu}\mathcal{K}^\rho_{\rho]}  , \quad
U_4  = \mathcal{K}^\mu_{[\mu}\mathcal{K}^\nu_{\nu}\mathcal{K}^\rho_{\rho}\mathcal{K}^\sigma_{\sigma]} ,
\end{align}
where the antisymmetrization is defined as usual (e.g.~$\mathcal{K}^\mu_{[\mu} \mathcal{K}^\nu_{\nu]} =(\mathcal{K}^\mu_{\mu} \mathcal{K}^\nu_{\nu} -\mathcal{K}^\mu_{\nu}\mathcal{K}^\nu_{\mu})/2$) and
\be
\mathcal{K}^\mu_\nu = \delta^\mu_\nu-\sqrt{g^{\mu\rho}f_{AB}\pd_\rho \phi^A \pd_\nu \phi^B }   .
\ee
$f_{AB}$ is a fiducial metric, which is often chosen as Minkowski $f_{AB}=\eta_{AB}$. We will for simplicity consider the Minkowski case when we study the cosmology of the model. The four $\phi^A(x)$ are St\"{u}ckelberg scalars introduced to restore general covariance \cite{ArkaniHamed:2002sp}, so that the theory (\ref{thefirstmodel}) is invariant under
\be
g_{\mu\nu}(x) \to \frac{\pd x^\rho}{\pd x'^\mu} \frac{\pd x^\sigma}{\pd x'^\nu} g_{\rho\sigma}(x), \quad \phi^A(x) \to \phi^A(x) ;\qquad x^\mu \to x'^{\mu} .
\ee
For the case $f_{AB}=\eta_{AB}$, the St\"{u}ckelberg scalars form Lorentz 4-vectors in the internal space (or, field space) labelled by $A, B, ...\;$. So in this case the theory also has the (global) Poincar\'e symmetry
\be
\phi^A \to \phi^A +c^A , \quad  \phi^A \to \Lambda^A{}_B\phi^B , \quad {\rm when}~~f_{AB}=\eta_{AB}  .
\ee
We may have a complete gauge fix by conveniently specifying the functional forms of $\phi^A(x)$. For example, we can choose ``unitary gauge'' $\phi^A=x^A$, leading to $\mathcal{K}^\mu_\nu = \delta^\mu_\nu-\sqrt{g^{\mu\rho}f_{\rho\nu}}$. We will work in this gauge to show that the theory is free of the BD ghost. Note that, even if $f_{AB}$ is not Minkowski, for some cases, we can choose a gauge so that $f_{AB}\pd_\rho \phi^A \pd_\nu \phi^B$ is Minkowski.

\section{Absence of the Boulware-Deser ghost} \label{absencebdghost}

In this section, we show that the promotion of the mass parameter to a function of a scalar does not spoil the absence of the Boulware-Deser ghost of the dRGT structure. We will first show that the Hamiltonian constraint is maintained in mass-varying massive gravity, and then prove the existence of a secondary constraint, arising from the time preservation of the primary Hamiltonian constraint, and compute the secondary constraint explicitly. We also argue that there is no tertiary constraint. To do these, it is more convenient to work in the following representation \cite{Hassan:2011vm}
\be \label{themodel}
S= \int \ud^4 x \mathcal{L}=\int \ud^4 x \sqrt{-g} \left[ \frac{M_P^2}{2} R + V(\psi) \left(\sum_{n=0}^{4}\beta_n e_n  \right)  -  \frac12 \pd_\mu \psi \pd^\mu \psi -W(\psi) \right]
\ee
where
\begin{align}
e_0  = 1 , \quad
e_1  = \mathcal{X}^\mu_\mu  ,\quad
e_2  =  \mathcal{X}^\mu_{[\mu}\mathcal{X}^\nu_{\nu]} ,\quad
e_3  = \mathcal{X}^\mu_{[\mu}\mathcal{X}^\nu_{\nu}\mathcal{X}^\rho_{\rho]}  ,\quad
e_4  = \mathcal{X}^\mu_{[\mu}\mathcal{X}^\nu_{\nu}\mathcal{X}^\rho_{\rho}\mathcal{X}^\sigma_{\sigma]}= \det(\mathcal{X}^\mu_\nu)
\end{align}
with
\be
\mathcal{X}^\mu_\nu \equiv \sqrt{g^{\mu\rho}f_{\rho\nu}},\quad \beta_n=(-1)^n\left( \frac12 (4-n)(3-n)+(4-n)\alpha_3+\alpha_4 \right)   .
\ee

First, we decompose the two metrics in the ADM formulation to identify the apparent dynamical degrees of freedom. Introducing the lapse functions, shift vectors and 3D induced metrics, we can rewrite the metrics as
\bea
g_{\mu\nu}\ud x^\mu \ud x^\nu &= - N^2\ud t^2 +\gamma_{ij}(\ud x^i+N^i\ud t)(\ud x^j+N^j\ud t) \\
f_{\mu\nu}\ud x^\mu \ud x^\nu &= - L^2\ud t^2 +\xi_{ij}(\ud x^i+L^i\ud t)(\ud x^j+L^j\ud t)  .
\eea
We will lower and raise the spatial indices by $\gamma_{ij}$ and its inverse $\gamma^{ij}$ respectively. The fiducial metric $f_{\mu\nu}$, the lapse $N$ and the shift $N_i$ do not have kinetic terms in the Lagrangian (\ref{themodel}), so they are non-dynamical. The potential dynamical degrees of freedom are $\gamma_{ij}$ and the scalar $\psi$. $\gamma_{ij}$ has 6 components, while a spin-2 field has only 5 propagating modes. If there were no constraints on the dynamical degrees of freedom arising from the equations of motion of $N$ or $N^i$ and their time conservation, then there would be six modes and the sixth mode would appear as the Boulware-Deser ghost  \cite{Creminelli:2005qk}. dRGT massive gravity, by carefully choosing the graviton potential, produces two constraints \cite{Hassan:2011hr, Hassan:2011ea}, eliminating the Boulware-Deser ghost. In the scalar-dRGT model (\ref{themodel}), as we will see,  two constraints also arise and the Boulware-Deser ghost can also be eliminated.

\subsection{The Hamiltonian constraint}

Defining the conjugate momenta
\be
\pi^{ij} = \frac{\pd\mathcal{L}}{\pd \dot{\gamma}_{ij}}, \qquad \pi = \frac{\pd\mathcal{L}}{\pd \dot{\psi}},
\ee
and performing the Legendre transformation, we can switch to the Hamiltonian formulation. In General Relativity, $N$ and $N^i$ appear as Lagrange multipliers in the Hamiltonian, so their equations of motion clearly impose constraints on the dynamical variables and their conjugate momenta. In the model (\ref{themodel}), none of the ADM variables $N$ and $N^i$ appears as a Lagrange multiplier, but the 4 equations of motion for $N$ and $N^i$ can actually give rise to an equation independent of $N$ and $N^i$,  i.e., a constraint on $\gamma_{ij}$, $\psi$ and their conjugate momenta. That is, there is an implicit Lagrange multiplier in the Hamiltonian. For the case at hand, we can make this Lagrange multiplier explicit by re-defining the shift vector $N^i$, and the Lagrange multiplier turns out to be $N$. To achieve this, we introduce a new shift vector $n^j$, following \cite{deRham:2010kj, Hassan:2011hr, Hassan:2011tf}
\be
N^i - L^i = (L\delta^i_j+ND^i{}_j)n^j  ,
\ee
where $D^i{}_j$ is defined by the following matrix equation
\be \label{Ddefrel}
x D^i{}_k D^k{}_j = (\gamma^{ik}-D^i{}_m n^m D^k{}_l n^l)\xi_{kj}, \quad x = 1-n^i\xi_{ij}n^j  .
\ee
The solution to this equation can be written in matrix form as
\be
D=(xI+nn^T\xi)^{-1}\sqrt{(xI+nn^T\xi)\Gamma^{-1}\xi}  ,
\ee
where the various matrices mean that $I=(\delta^i_j)$, $\xi=(\xi_{ij})$ and $\Gamma=(\gamma_{ij})$. With this change of variables, after some algebra, the Hamiltonian can be written as
\be \label{hamil}
H=\int \ud x^3 \mathcal{H} =\int \ud x^3 (\mathcal{H}_0  - N\mathcal{C})
\ee
where
\begin{align}
\mathcal{H}_0 &= -(Ln^i+L^i)\mathcal{R}_i-L\sqrt{\gamma}\mathcal{A}V(\psi)  , \\
\mathcal{C}  &= \mathcal{R}+\mathcal{R}_i D^i{}_jn^j+ \sqrt{\gamma}\mathcal{B} V(\psi).
\end{align}
Here $\mathcal{R}$, $\mathcal{R}_i$, $\mathcal{A}$ and $\mathcal{B}$ are given by
\begin{align}
\mathcal{R} &=\frac{\sqrt{\gamma}}{2}(M_P^2{\,}^{(3)}\!R-\pd_i\psi\pd^i\psi-2W(\psi))+\frac{1}{\sqrt{\gamma}} (\frac{1}{M_P^2} \pi^i{}_i\pi^j{}_j-\frac{2}{M_P^2} \pi^{ij}\pi_{ij} - \frac12 \pi^2)  ,\\
\mathcal{R}_i &= 2\gamma_{ij} \nabla_k \pi^{jk} -\pi\pd_i \psi   , \\
\mathcal{A} & = \beta_1 x^{\frac12}+\beta_2[ xD^i{}_i+n^i\xi_{ij}D^j{}_kn^k ]
+\beta_3 \left[2x^{\frac12}D^{[l}{}_ln^{i]}\xi_{ij}D^j{}_kn^k+x^{\frac32}D^{[i}{}_iD^{j]}{}_j\right] + \beta_4\sqrt{\frac{\xi}{\gamma}}  , \\
\mathcal{B} & = \beta_0 + \beta_1 x^{\frac12}D^i{}_i + \beta_2 x D^{[i}{}_iD^{j]}{}_j
          +\beta_3 x^{\frac32} D^{[i}{}_iD^j{}_jD^{k]}{}_k    .
\end{align}
Now, $n^i$ are auxiliary fields and can in principle be integrated out by imposing their equations of motion $\pd \mathcal{L}/\pd n^i = -\pd \mathcal{H}/\pd n^i=0$, or equivalently imposing the following three equations
\begin{align}
\mathcal{C}_i &= \mathcal{R}_i -V(\psi)\sqrt{\gamma}x^{-\frac12}n^l\xi_{lj}\left[\beta_1\delta^j_i+\beta_2 x^{-\frac12}(\delta^j_i D^k{}_k-D^j{}_i) +\beta_3 x(\delta^j_i D^{[k}{}_k D^{l]}{}_l - 2D^{[k}{}_k D^{j]}{}_i )  \right] \nn
   &  =  0 .
\end{align}
After this, the equation of motion for $N$, which is a Lagrange multiplier in the Hamiltonian (\ref{hamil}), becomes a constraint
\be
\mathcal{C}(\gamma_{ij},\pi^{ij},\psi,\pi,n^i(r_{ij},\pi^{ij},\psi,\pi)) =  0 .
\ee
This is the Hamiltonian constraint.

\subsection{Existence of a secondary constraint}

Now, a constraint should be preserved under the evolution of the dynamical system, so for consistency the following equation should be satisfied
\be \label{conspers}
\frac{\ud}{\ud t}\mathcal{C}(x) = \{\mathcal{C}(x),H\}= \int \ud^3 y \left[\{\mathcal{C}(x),\mathcal{H}_0(y)\}- N(y)\{\mathcal{C}(x),\mathcal{C}(y)\}\right]=0  ,
\ee
where $\{,\}$ is the Poisson bracket, defined by
\begin{align} \label{poissonb}
\{A(x), B(y) \} &= \int \ud^3 z \left( \frac{\delta A(x)}{\delta \gamma_{mn}(z)} \frac{\delta B(y)}{\delta \pi^{mn}(z)}
-\frac{\delta A(x)}{\delta \pi^{mn}(z)} \frac{\delta B(y)}{\delta \gamma_{mn}(z)} \right) \nn
    &~~~+\int \ud^3 z \left( \frac{\delta A(x)}{\delta \psi(z)} \frac{\delta B(y)}{\delta \pi(z)}
-\frac{\delta A(x)}{\delta \pi(z)} \frac{\delta B(y)}{\delta \psi(z)} \right)  .
\end{align}

If the equation (\ref{conspers}) is satisfied trivially, our constraint system is self-consistent and no further checking is needed. If not, we should check whether this equation determines the lapse $N$. We can see that if $\{\mathcal{C}(x),\mathcal{C}(y)\}$ does not vanish on the primary constraint surface $\mathcal{C}=0$, then the equation (\ref{conspers}) can be used to determine $N$ and no secondary constraint arises. If the equation (\ref{conspers}) can not be used to determine $N$ (and is not trivially satisfied), a secondary constraint arises. That is, if $\{\mathcal{C}(x),\mathcal{C}(y)\}$ vanishes upon using $\mathcal{C}=0$, we will have a secondary constraint
\be
\mathcal{C}^{(2)}(x) = \int \ud^3 y \{\mathcal{C}(x),\mathcal{H}_0(y)\} =0 .
\ee
In the following, we will show that $\{\mathcal{C}(x),\mathcal{C}(y)\}$ does vanish upon using $\mathcal{C}=0$.

From Eq.~(\ref{Ddefrel}) and $n^i$'s equation of motion, we can infer $\pd \mathcal{C}/\pd n^i = 0$, which means we have
\be
\delta \mathcal{C} = \left.\frac{\pd \mathcal{C}}{\pd \gamma_{mn}}\right|_{n^i} \delta \gamma_{mn} + \left.\frac{\pd \mathcal{C}}{\pd \pi^{mn}}\right|_{n^i}  \delta \pi^{mn} +\left.\frac{\pd \mathcal{C}}{\pd \psi}\right|_{n^i} \delta \psi + \left.\frac{\pd \mathcal{C}}{\pd \pi}\right|_{n^i}  \delta \pi    .
\ee
Now, we can expand
\begin{align}
\{ \mathcal{C}(x), \mathcal{C}(y) \} & = \{\mathcal{R}(x), \mathcal{R}(y)\} + \{\mathcal{R}_i(x), \mathcal{R}_j(y)\} D^i{}_k n^k(x) D^j{}_ln^l(y) \nn
  &~~~+\left[ \{\mathcal{R}(x), \mathcal{R}_i(y)\} D^i{}_k n^k(y)
  + S^{mn}(x)\frac{\delta \mathcal{R}_i(y)}{\delta \pi^{mn}(x)} D^i{}_kn^k(y)  - (x\leftrightarrow y)\right]  ,
\end{align}
where
\begin{align} \label{smn}
S^{mn}(x) = \mathcal{R}_j\frac{\pd (D^j{}_r n^r)}{\pd \gamma_{mn}}(x)+V(\psi)\frac{\pd (\sqrt{\gamma}\mathcal{B})}{\pd \gamma_{mn}}(x)  .
\end{align}
In this expansion, we have dropped terms that are proportional to $\delta^3(x-y)$, since $\{ \mathcal{C}(x), \mathcal{C}(y) \}$ is antisymmetrical in exchanging $x$ and $y$. In particular, the terms involving the factor $\pd V(\psi)/\pd \psi$ cancel each other.

The Poisson brackets of $\mathcal{R}$ and $\mathcal{R}_i$ satisfy the same algebra as that of pure General Relativity:
\begin{align} \label{grsalg}
\{\mathcal{R}(x), \mathcal{R}(y)\}  &=-\left[\mathcal{R}^i(x)\pd_{x^i} \delta^3(x-y) - (x\leftrightarrow y)\right]  ,\nn
\{\mathcal{R}(x), \mathcal{R}_i(y)\}&=-\mathcal{R}(y)\pd_{x^i} \delta^3(x-y)  ,  \nn
\{\mathcal{R}_j(x), \mathcal{R}_j(y)\} &=-\left[\mathcal{R}_j(x)\pd_{x^i} \delta^3(x-y) - (x\leftrightarrow y)\right]  .
\end{align}
To show this, one can simply expand the Poisson brackets with
\begin{align}
\mathcal{R}  &= {\mathcal{R}}^{GR}-\frac{\sqrt{\gamma}}{2}(\pd_i\psi\pd^i\psi+2W(\psi))-\frac{1}{2\sqrt{\gamma}} \pi^2 ,\\
\mathcal{R}_i &= {\mathcal{R}}_i^{GR}  -\pi\pd_i \psi    ,
\end{align}
utilize the algebra of pure General Relativity \cite{Khoury:2011ay}, and the smoothing function technique introduced below.

To deal with the Dirac delta function's derivatives, we introduce smoothing functions $f(x)$ and $g(x)$,  and the quantities
\be
F = \int \ud^3 x f(x) \mathcal{C}(x) ,\qquad   G = \int \ud^3 y g(y) \mathcal{C}(y) .
\ee
We require that $f(x)$ and $g(x)$ fall off quickly enough at infinities so that boundary terms can be dropped when integrating by parts.

So we shall first calculate the Poisson bracket
\be \label{fgdef}
\{ F, G \} = \int \ud^3 x \ud^3 y f(x) g(y) \{ \mathcal{C}(x), \mathcal{C}(y) \}   .
\ee
After using the algebra (\ref{grsalg}), some partial integration and carrying out one of the integrals, we get
\be \label{fgp}
\{F,G\} = -\int d^3x \Big(f \, \pd_i g-g\, \pd_i f\Big)\, \mathcal{P}^i\,,
\ee
where
\be
\mathcal{P}^i=(\mathcal{R} + \mathcal{R}_j D^j_{~k}n^k) D^i_{~l}n^l +\mathcal{R}^i+  S^{il}\gamma_{lj}
D^j_{~k}n^k \, .
\ee
Then $\{ \mathcal{C}(x), \mathcal{C}(y) \}$ can be extracted by writing $f(x)=\int \ud^3 y f(y) \delta^3(x-y)$ and $g(y)=\int \ud^3 x g(x) \delta^3(x-y)$ and comparing Eq.~(\ref{fgp}) to Eq.~(\ref{fgdef}), which gives
\be
\{{\cal C}(x),{\cal C}(y) \}= - \left[
\mathcal{P}^i(x)\, \pd_{x^i} \delta^3(x-y)
-(x \leftrightarrow  y) \right] \, .
\ee

To proceed, we have to calculate the partial derivatives $\pd (D^j{}_r n^r)/\pd \gamma_{mn}$ and $\pd (\sqrt{\gamma}\mathcal{B})/\pd \gamma_{mn}$ in $S^{mn}$, using the defining relation of $D^i{}_k$, i.e., Eq.~(\ref{Ddefrel}). For this, we can essentially follow the dRGT massive gravity case \cite{Hassan:2011hr}, as these partial derivations do not involve differentiation with respect to $\psi$. After some lengthy calculation, we get
\be
\mathcal{P}^i  = \mathcal{C} D^i{}_l n^l  .
\ee
So $\{ \mathcal{C}(x), \mathcal{C}(y) \}$ vanishes on the surface of $\mathcal{C}=0$.

\subsection{The secondary constraint}

Now, we shall derive the expression for the secondary constraint
\be
\mathcal{C}^{(2)}(x) = \int \ud^3 y \{\mathcal{C}(x),\mathcal{H}_0(y)\} .
\ee
To this end, we can expand the Poisson bracket
\begin{align}
\!\{{\cal C}(x) , {\cal H}_0(y)\}&
\!=\!-\{\mathcal{R}(x),\mathcal{R}_i(y)\}(Ln^i\!+\!L^i)(y)-D^i_{~k}n^k(x)\,
\{\mathcal{R}_i(x),\mathcal{R}_j(y)\}(Ln^j\!+\!L^j)(y) \nn
&~~+L\frac{\delta \mathcal{R}(x)}{\delta\pi^{mn}(y)}\,\sqrt{\gamma}V(\psi)A^{mn}(y)
+ L\,D^i_{~k}n^k(x)\,\frac{\delta \mathcal{R}_i(x)}{\delta\pi^{mn}(y)}
\sqrt{\gamma}V(\psi)A^{mn}(y)\nn
&~~-S^{mn}(x)\,\frac{\delta \mathcal{R}_i(y)}{\delta\pi^{mn}(x)}\,(Ln^i\!+\!L^i)(y)
- \sqrt{\gamma} \mathcal{B} \frac{\pd V}{\pd \psi}\frac{\pd\mathcal{R}_i}{\pd\pi} (Ln^i+L^i)\delta^3(x-y) \nn
&~~+L\sqrt{\gamma} \mathcal{A} \frac{\pd V}{\pd \psi}\frac{\pd\mathcal{R}}{\pd\pi} \delta^3(x-y) +L \sqrt{\gamma} \mathcal{A}  \frac{\pd V}{\pd \psi}\frac{\pd\mathcal{R}_i}{\pd\pi}D^i{}_kn^k \delta^3(x-y)
\end{align}
where $S^{mn}(x)$ is defined as Eq.~(\ref{smn}) and $A^{mn}$ is defined as
\be
A^{mn}\equiv \frac{1}{\sqrt{ \gamma}}\,
\frac{\pd\left(\sqrt{ \gamma}\mathcal{A}\right)}{\pd\gamma_{mn}}\, .
\ee
Note that we have exploited the relations  $\pd \mathcal{C}/\pd n^i=0$ and $\pd \mathcal{H}_0/\pd n^i=0$.
After some integration by parts and using the Poisson bracket algebra for $\mathcal{R}$ and $\mathcal{R}_i$ and Eq.~(\ref{fgdef}), the secondary constraint can be written as
\begin{align}
\mathcal{C}^{(2)} & ={\cal C}  \nabla_i(Ln^i+L^i) +
V(\psi) L \left(\gamma_{mn}\pi^k{}_k-2\pi_{mn}\right)A^{mn}
+2L\sqrt{\gamma} \nabla_m (V(\psi) A^{mn})\gamma_{ni} D^i_{~k} n^k
\nn
&~~+\left(R_j D^i_{~k}n^k -V(\psi)\sqrt{ \gamma}\gamma_{jk}\bar{\mathcal{B}}^{ki}
\right) \nabla_i(Ln^j+L^j)
+(\nabla_i \mathcal{R} + \nabla_i \mathcal{R}_jD^j_{~k}n^k)(Ln^i+L^i)
\nn
 &~~   - \frac{\pd V}{\pd \psi}\left[\sqrt{\gamma} \mathcal{B}\pd_i \psi (Ln^i+L^i)
+L\mathcal{A} \pi + L\sqrt{\gamma} \mathcal{A} \pd_i \psi  D^i{}_k n^k \right] \nn
&=0
\end{align}
where
\begin{align}
\bar{\mathcal{B}}^{mn}\equiv &\gamma^{mi}\Bigg[\beta_1x^{-\frac12}
  \xi_{ik}  (D^{-1})^k_{~j} + \beta_2\left(\xi_{ik}
  (D^{-1})^k_{~j}D^l_{~l}-\xi_{ij}\right) \nn
&+ \beta_3  x^{\frac12}  \left(\xi_{ik}  D^k_{~j}- \xi_{ij}
  D^k_{~k} +\frac{1}{2} \xi_{ik}  (D^{-1})^k_{~j} (D^l_{~l}
  D^h_{~h}-D^l_{~h}D^h_{~l})\right) \Bigg] \,\gamma^{jn}\,.
\end{align}

\subsection{Absence of a tertiary constraint}

We also have to check whether the time preservation of $\mathcal{C}^{(2)}$,
\be
\frac{\ud}{\ud t}\mathcal{C}^{(2)}(x) = \{\mathcal{C}^{(2)}(x),H\} = 0 ,
\ee
can give rise to a tertiary constraint. One has to go through a discussion similar to that below Eq.~(\ref{poissonb}). But it is easy to see that there is no tertiary constraint.  In dRGT gravity, the corresponding $\{ \mathcal{C}^{(2)},\mathcal{H}_0 \}$ and $\{ \mathcal{C}^{(2)},\mathcal{C} \}$ do not vanish on the constraint surface. As dRGT gravity is a limit of mass-varying massive gravity, $\{ \mathcal{C}^{(2)},\mathcal{H}_0 \}$ and $\{ \mathcal{C}^{(2)},\mathcal{C} \}$ in mass-varying massive gravity also do not vanish on the constraint surface. Therefore the time preservation of $\mathcal{C}^{(2)}$ simply determines the lapse $N$ and no tertiary constraint arises.

\section{FRW Cosmology}   \label{frwcosmology}

In this section, we shall study the cosmology of mass-varying massive gravity (\ref{thefirstmodel}) in the homogeneous and isotropic background. As mentioned previous, we will choose, for simplicity, the fiducial metric to be Minkowski
\be
f_{AB}=\eta_{AB}
\ee
for considering the cosmology of the model. Note that this also covers the cases where $f_{AB}$ can be brought to the Minkowski metric by general coordinate transformation, as we can always choose a gauge for the St\"{u}ckelberg fields $\phi^A$. For an interesting cosmology with the correct hot history, we shall add some matter source
\be
S_T = S + S_m  , \quad -\frac{2}{\sqrt{-g}}\frac{\delta S_m}{\delta g^{\mu\rho}}g^{\rho\nu} = {\rm diag}(\rho_m, p_m, p_m, p_m)
\ee
which have been assumed to be a perfect fluid. For a generic FRW solution, the dynamical metric and the fiducial metric do not have to be diagonalized simultaneously \cite{Volkov:2012cf}. In this paper, however, we shall focus on the diagonal case, already with rich cosmologies, leaving the non-diagonal case for future work.

We shall discuss different FRW universes separately. Since our fiducial metric is Minkowski, whether we can have simultaneously diagonal metrics  for different spatial geometries depends on whether the Minkowski metric can be charted by these geometries. The flat case can be charted trivially by choosing ``unitary gauge'' $\phi^A=x^A$. While the open universe can still be charted, the closed universe does not admit a FRW chart  \cite{Gumrukcuoglu:2011ew}.

\subsection{Equations of motion}

\subsubsection{Flat universe}

For this case, the metric and St\"{u}ckelberg fields ansatz we are interested in can be written as
\begin{align}
\ud s^2 &= -N(\tau)^2 \ud \tau^2 +a(\tau)^2 \delta_{ij} \ud x^i \ud x^j  ,\quad~~  \phi^0 = b(\tau)  ,  ~~~~~\phi^i = a_0 x^i  ,
\end{align}
where $a_0$ is constant. Without lost of generality, we may assume that $a>0, a_0>0 , b'>0,  N>0$, the action for this background reduces to
\be
S_T = \int \ud^4 x \left[  -3M_P^2  \frac{a'^2a}{N} + V(\psi) (u_2^F +\alpha_3 u_3^F + \alpha_4 u_4^F)  +\frac{a^3}{2N} \psi'^2 -Na^3W(\psi)\right]   + S_m   ,
\ee
where
\begin{align}
u_2^F & = 3a(a-a_0)(2Na-a_0 N-ab')  ,  \\
u_3^F & = (a-a_0)^2 (4Na-a_0 N-3ab')  ,  \\
u_4^F & = (a-a_0)^3(N-b')    ,
\end{align}
and we have define ${~~}'=\ud /\ud \tau$ . If we also define
\be
X=\frac{a_0}{a}, \quad \dot{a}=\frac{\ud a}{ N\ud \tau}, \quad H=\frac{\dot{a}}{ a }    ,
\ee
and the polynomials
\begin{align}
f_1(\alpha_i,X) &= (3-2X)+\alpha_3(3-X)(1-X)+\alpha_4(1-X)^2 , \\
f_2(\alpha_i,X) &= (1-X)+\alpha_3(1-X)^2+\frac{\alpha_4}{3}(1-X)^3 , \\
f_3(\alpha_i,X) &=(3-X)+\alpha_3(1-X)  ,
\end{align}
then the equations of motion for $b$, $N$, $a$ and  $\psi$ are given by, respectively,
\begin{align}
\label{flateom1}
V(\psi)Hf_1(\alpha_i,X)+&\dot{V}(\psi)f_2(\alpha_i,X)  = 0   ,\\
\label{flateom2}
3M_P^2 H^2 &= \rho_{MG}+\rho_m     ,\\
\label{flateom3}
-2 M_P^2 \dot{H} &=\rho_{MG}+p_{MG}+\rho_m +p_m   ,\\
\label{flateom4}
\ddot{\psi}+3H\dot{\psi}+\frac{\ud W}{\ud \psi}+\frac{\ud V}{\ud \psi}[(X-1) & (f_3(\alpha_i,X)+f_1(\alpha_i,X))+3\dot{b}f_2(\alpha_i,X)]  =0 ,
\end{align}
where  $\rho_m$ and $p_m$ are the density and pressure for the matter source respectively and the effective density and pressure are given by
\begin{align}
\rho_{MG}& =\frac12 \dot{\psi}^2+W(\psi)+V(\psi) (X-1)f_3(\alpha_i,X)+V(\psi) (X-1)f_1(\alpha_i,X) , \\
p_{MG} & =\frac12 \dot{\psi}^2-W(\psi)- V(\psi)(X-1)f_3(\alpha_i,X)  -V(\psi)(\dot{b}-1) f_1(\alpha_i,X)  .
\end{align}
Note that $f_1(\alpha_i, X)$ and  $f_2(\alpha_i, X)$ satisfy a differential relation
\be
f_1(\alpha_i, X)= \frac{\ud [3 X^{-3} f_2(\alpha_i, X)]}{\ud (X^{-3})} = 3f_2(\alpha_i, X)-\frac{\ud f_2(\alpha_i, X)}{\ud \ln X}   .
\ee
Eq.~(\ref{flateom1}) can be integrated by the method of viable separation, which reduces to an algebraic equation for $V(\psi)$ and $a$,
\be
V(\psi) = 3V_0e^{-\int \frac{f_1}{af_2}\ud a}=\frac{a_0^3 V_0}{a^3f_2(\alpha_i, X)} ,\quad V_0 = {\rm const.}>0.
\ee
That is, this equation can be seen as a constraint\;\footnote{We thank Emmanuel Saridakis for pointing out this constraint equation.}.

Unlike the diagonal case of dRGT massive gravity, the system (\ref{flateom1}-\ref{flateom4}) can have interesting expanding solutions. But the system is highly non-linear, and, apart from some simple ones, exact solutions are difficult to find. One good strategy might be to apply a dynamical system analysis on it to extract its phase space information (see e.g.~\cite{Zhou:2009cy}). For this, one shall first reduce it to a standard dynamical system and analyze the equilibria the system possesses, which we leave for future work.

\subsubsection{Open universe}

For an open FRW universe, $K<0$ (we shall use the constant $k = \sqrt{|K|}$), the metric and St\"{u}ckelberg fields ansatz we are interested in can be written as
\begin{align}
\ud s^2 &= -N(\tau)^2 \ud \tau^2 +a(\tau)^2 \Omega_{ij} \ud x^i \ud x^j ,
\quad\Omega_{ij} \ud x^i \ud x^j = \delta_{ij} \ud x^i \ud x^j  -\frac{k^2(\delta_{ij} x^i \ud x^j)^2}{1+k^2(\delta_{ij} x^i x^j)} ,\\
\phi^0 &= b(\tau)  \sqrt{1+k^2(\delta_{ij} x^i x^j)}, ~~~~~~~\phi^i = k b(\tau)x^i  .
\end{align}
For this ansatz, $g_{\mu\nu}$ and $\eta_{AB}\pd_\mu \phi^A\pd_\nu \phi^B$ become diagonal simultaneously, so $\mathcal{K}^\mu_\nu$ becomes diagonal and its square root is uniquely defined and easy to calculate. Assume that $a>0, b>0, b'>0,  N>0$, the action now reduces to
\begin{align}
S_T = \int \ud^4 x \sqrt{\Omega}\Big[   -3M_P^2( k^2 N a + &\frac{a'^2a}{N}) + V(\psi) (u_2^O +\alpha_3 u_3^O + \alpha_4 u_4^O)  \nn
 &  +\frac{a^3}{2N} \psi'^2-Na^3W(\psi)\Big]  +S_m,
\end{align}
where
\begin{align}
u_2^O & = 3a(a-kb)(2Na-kNb-ab')    \\
u_3^O & = (a-kb)^2 (4Na-kNb-3ab')  \\
u_4^O & = (a-kb)^3(N-b')
\end{align}
Defining
\be
Y=\frac{kb}{a}, \quad \dot{a}=\frac{\ud a}{ N\ud \tau}, \quad H=\frac{\dot{a}}{ a }
\ee
the equations of motion for $b$, $N$, $a$ and  $\psi$ are given by, respectively,
\begin{align}
\label{openeom1}
V(\psi)\left(H-\frac{k}{a}\right)f_1(\alpha_i,Y)&+\dot{V}(\psi)f_2(\alpha_i,Y)  = 0   ,\\
\label{openeom2}
M_P^2(3H^2-\frac{3k^2}{a^2})& = \rho_{MG}+\rho_m     ,\\
\label{openeom3}
M_P^2(-2\dot{H}-2\frac{k^2}{a^2})=&\rho_{MG}+p_{MG}+\rho_m +p_m   ,\\
\label{openeom4}
\ddot{\psi}+3H\dot{\psi}+\frac{\ud W}{\ud \psi}+\frac{\ud V}{\ud \psi}[(Y-1)(f_3(\alpha_i,Y)&+f_1(\alpha_i,Y))+3\dot{b}f_2(\alpha_i,Y)]  =0 ,
\end{align}
where the polynomials $f_1(\alpha_i, Y)$, $f_2(\alpha_i, Y)$ and $f_3(\alpha_i, Y)$ are defined the same as that in the flat universe case,  $\rho_m$ and $p_m$ are the density and pressure for the matter source respectively, and the effective density and pressure are given by
\begin{align}
\rho_{MG}& =\frac12 \dot{\psi}^2+W(\psi)+V(\psi) (Y-1)f_3(\alpha_i,Y)+ V(\psi) (Y-1)f_1(\alpha_i,Y) , \\
p_{MG} & =\frac12 \dot{\psi}^2-W(\psi)- V(\psi) (Y-1)f_3(\alpha_i,Y)  -V(\psi)(\dot{b}-1) f_1(\alpha_i,Y) .
\end{align}
Again, in general, the system (\ref{openeom1}-\ref{openeom4}) is highly non-linear, and, exact cosmological solutions are difficult to obtain and one may apply a dynamical system analysis on it to extract its phase space information.

\subsection{A special solution and its cosmological implications}   \label{cosmoscena}

In the open universe case, there exists one simple branch of solutions if we have
\be \label{a34}
\alpha_4 =\frac{ 3 \alpha_3^2}{4},  \qquad Y = \frac{kb}{a} = \frac{2 + \alpha_3}{\alpha_3} ,
\ee
that is, if we impose a constraint on the free parameters $\alpha_3$ and $\alpha_4$ and let the two scale factors $b(t)$ and $a(t)$ scale with each other. This gives rise to
\be
f_1(\alpha_i, Y)=f_2(\alpha_i, Y)=0 .
\ee
In this branch, the dynamical system (\ref{openeom1}-\ref{openeom4}) reduces to that of a General Relativity plus a scalar
\begin{align}
M_P^2(3H^2-\frac{3k^2}{a^2})& = \frac12 \dot{\psi}^2+\bar{W}(\psi)+\rho_m     ,\\
M_P^2(-2\dot{H}-2\frac{k^2}{a^2})&=\dot{\psi}^2+\rho_m +p_m   ,\\
\ddot{\psi}+3H\dot{\psi}+\frac{\ud \bar{W}}{\ud \psi} &=0 .
\end{align}
where
\be
\bar{W}(\psi)=W(\psi)-\frac{4}{\alpha_3^2}V(\psi)   .
\ee
The evolution of this system will determine the effective mass $\sqrt{V(\psi)/M_P^2}$ of the graviton.

As an aside, we could have for simplicity set $W(\psi)$ to 0 so that the scalar potential $\bar{W}(\psi)$ would be simply supported by $V(\psi)$. For the graviton sector to be free of tachyonic instabilities, we need $V(\psi)>0$, which leads to $\bar{W}(\psi)<0$, that is, the scalar system can only have negative potential energy (we can still require it to be bounded below though.). In this case, the graviton mass could only increase and we would not have a de Sitter-like solution but have an anti-de Sitter-like solution instead.

In this special case, the background evolution of mass-varying massive gravity connects to some well studied cosmological systems. We may identity this scalar with inflaton, quintessence, quintessential-inflaton, or any other scalar that plays a role in the homogeneous and isotropic evolution of the universe. In constructing these scenarios, however, one shall make sure the effective graviton mass in the current cosmic time is small enough to satisfy all the existing General Relativity tests.

For example, here we discuss a simple model that gives rise to inflation and dark energy. We shall consider
\be
W(\psi)= M^2 \psi^2  ,\qquad V(\psi) =  \frac12 \alpha_3^2 m M^2 (\psi - m)   ,
\ee
which leads to an effective potential $\bar{W}=M^2(\psi-m)^2+m^2M^2$, where $m$ and $M$ are free parameters of the model, to be adjusted to satisfy observations. To give a concrete example, if we choose $M\sim 10^{-5}M_P$ and $m\sim 10^{5} H_0$ ($H_0$ is the current Hubble constant), and set the $\psi$ field initially to be around the Planck scale $M_P$, then the universe will start with chaotic inflation, reheat around the minimum of $\bar{W}$ at $\psi=m\sim 10^{5} H_0$ and has, at the background level, an effective cosmological constant $m^2M^2\sim H_0^2M_P^2$, which gives rise to dark energy. Interestingly, around this vev, the graviton is massless and its potential starts only from the cubic order in perturbation. One may further consider whether this model will reproduce all the successes of inflation's power spectra. For this one shall study the perturbations around the quasi-de Sitter background generated by $\bar{W}(\psi)$. Here we give a simple argument which implies that these successes might be reproduced. In massive gravity, there is a Vainshtein radius for some matter source, within which massive gravity effectively reduces to Einstein gravity.  In the early universe the squared mass of the graviton is $\sim\alpha_3^2 m M^2 \psi/M_P^2$.  For a spherical source $\mathcal{M}$ the Vainshtein radius for  mass-varying massive gravity is given by $(\mathcal{M}/m_g^2M_P^2)^{1/3}$, where $m_g$ is the graviton mass. We may see a Hubble patch as a spherical source $\mathcal{M}\sim\rho_T H^{-3}\sim M^2_PH^2\cdot H^{-3}$ ($H^{-1}$ is the Hubble radius) and estimate the Vainshtein radius as
\be
r_V=(\alpha_3^2 H_0H^2)^{-\frac13} .
\ee
Even if the free parameter $\alpha_3$ is chosen as $\mathcal{O}(1)$, the Vainshtein radius $r_V$ is still much greater than the Hubble radius at that time. Therefore we may expect that all the success of the conventional inflation model will still hold. However, this should be confirmed by an explicitly perturbation analysis, which might also give rise to some testable differentiating effects. We leave this for future work. After reheating the inflaton settles down on the vev of the effective potential, which is of the dark energy scale, and the graviton becomes massless, so the big bang hot history is recovered afterwards.

\section{Discussions}  \label{conclusion}

Particle masses can often be generated by a scalar field, like Higgs. In this paper, we have introduced a model where the graviton mass is effectively generated by a scalar field, which spontaneously breaks general covariance of General Relativity. This model is a generalization of dRGT massive gravity and is also free of the BD ghost. We have proven the absence of the BD ghost by explicitly obtaining two constraints in the Hamiltonian analysis of the theory.

We may further generalize our model by promoting $\alpha_3$ and $\alpha_4$ to functions of $\psi$, i.e., $\alpha_3(\psi)$ and $\alpha_4(\psi)$. It is straightforward to check that a similar proof can go through, thus the BD ghost is not re-introduced either. We may also consider a non-minimal coupling $\sqrt{-g}h(\psi)R$, $h(\psi)$ being a general function of $\psi$. For this case, we can do a conformal transformation, which reduces the non-minimal coupling to the Einstein case but introduces some non-canonical kinetic terms and scale the graviton potential terms with powers of $h(\psi)$. We also expect this generalization is free of the BD ghost. Further more, we can make the fiducial metric a genuine dynamical one \cite{Hassan:2011ea}, which should also not spoil the absence of the BD ghost.

We have also initiated the study of the homogenous and isotropic cosmology for this model. We have derived the background equations of motion for the case when the two metrics can be diagonalized simultaneously, pending a detailed analysis of the systems of differential equations and the non-diagonal case for future work. Nevertheless, we have found a simple solution when $\alpha_3$ and $\alpha_4$ satisfy the relation (\ref{a34}). In this special case, the model's background evolution completely reduces to that of General Relativity plus a scalar. We have tentatively considered a simple scenario where the mass-varying scalar is identified with the chaotic inflaton and its low energy vev gives rise to the current cosmic acceleration.

In that case, the graviton mass is initially relatively large and settles to zero at reheating. Of course, one may alternatively consider various other scalar potentials, many of which might also lead to interesting cosmologies. If we draw more direct comparison with the spontaneous symmetry breaking in the Standard Model, one may consider a scalar potential where the graviton is initially massless and gains a small mass after symmetry breaking. Then, an interesting question may be: what is the relation between this scalar and the Higgs boson (or the Higgs bosons)? A scalar coupled to gravity might respond substantially to some local matter distribution, which might induce spatial variation of the graviton mass. As we generally require the current graviton mass is small, the spatial variation effect may arguably be small, but it is interesting to understand whether the current experiments and observations can impose any theoretical bounds on mass-varying massive gravity from this perspective.

~\\

{\bf Acknowledgements} We thank Claudia de Rham, Andrew Tolley, Antonio Padilla, Paul Saffin and Emmanuel Saridakis for helpful discussions. QGH is supported by the project of Knowledge Innovation Program of Chinese Academy of Sciences and a grant from NSFC (grant NO. 10975167). YSP is supported in part by NSFC
under Grant No:11075205, in part by the Scientific Research Fund
of GUCAS(NO:055101BM03), in part by National Basic Research
Program of China, No:2010CB832804. SYZ acknowledges hospitality
and financial support from the Institute of Theoretical Physics
(CAS, Beijing) during completion of this work.

\end{document}